\documentclass[prb]{revtex4}%
\usepackage{graphicx}
\usepackage{amsmath}
\usepackage{amsfonts}
\usepackage{amssymb}%
\providecommand{\U}[1]{\protect\rule{.1in}{.1in}}
\begin{document}
\author{Norman H. March$^{1,2,3}$, Alejandro Cabo$^{1,4}$ and Francisco Claro$^{5}$ }
\title{Phase diagram of two dimensional electron gas in a perpendicular magnetic
field around Landau level filling factors $\nu=1$ and $3$. }
\affiliation{$^{1}$Abdus Salam International Center for
Theoretical Physics, Trieste, Italy.} \affiliation{$^{2}$
Department of Physics, University of Antwerp, Antwerp, Belgium.}
\affiliation{$^{3}$ Oxford University, Oxford, England.}
\affiliation{$^{4}$ Grupo de F\'{\i}sica Te\'{o}rica, Instituto de
Cibern\'{e}tica, Matem\'atica y F\'{\i}sica, La Habana, Cuba}
\affiliation{$^{5}$ Facultad de F\'{\i}sica, Pontificia
Universidad Cat\'{o}lica de Chile, Santiago de Chile, Chile.}

\begin{abstract}
The measured melting curve $T_{m}(\nu)$ between the crystal and liquid phases
is analyzed using thermodynamics to extract the change of magnetization
$\Delta M$ as a function of the Landau level filling factor \ $\nu,$ near
$\nu=1$. \ An explanation of $\Delta M$($\nu)$ is proposed \ in terms of
Skyrmions. \ Near $\nu=3$, a Wigner crystal is the most probable solid phase,
experiments excluding Skyrmions.

\end{abstract}
\maketitle

\bigskip



\ In previous works \cite{cabclarmarch, cabclarmarch1, cabclar} the authors
have proposed the possible existence of \ a Hall crystal and/or Skyrmions at
very low temperatures near the Landau filling factor $\nu=1.$ The recent $NMR$
\ relaxation experiments of Gervais $et$ $al$ \cite{gervais} strengthen the
earlier claim of Barrett $et$ $al$ \cite{barret} that Skyrmions exist around
$T\sim40$ $mK$ near $\nu=1$, and a possible transition from a liquid to a
lattice of \ Skyrmions could occur there. We point out here the
thermodynamical result \cite{leamarch} for the slope of the melting curve
$T_{m}(\nu)$:
\begin{equation}
\frac{\partial}{\partial\nu}\ T_{m}(\nu)=\frac{B}{\nu}\frac{\Delta M}{\Delta
S}, \label{marchlea}%
\end{equation}
\begin{figure}[h]
\includegraphics[width=3.0in]{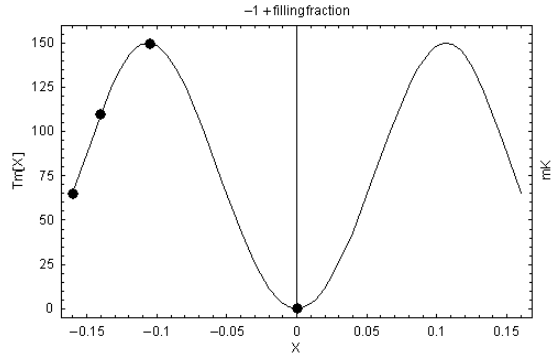}\caption{The experimental measurements
of Gervais at al for the solid to liquid transition temperatures $T_{m}$ for
$\nu=0.84,0.86,0.895$ are shown as black points. The solid curve is a fit to
an even polynomial curve of $X=\nu-1$, which is also satisfying $T_{m}=0$ at
$X=0$ as required by the zero density of Skyrmions at $\nu=1.$}%
\label{march}%
\end{figure}where $\Delta M$ and $\Delta S$ represent magnetization and
entropy changes respectively at the transition temperature $T_{m}$
of Skyrmion melting and $B$ and $\nu$ are the magnetic field and
the filling factor, respectively. The three estimated values of \
the solid-liquid transition temperature $T_{m},$ according to the
measurements of Gervais $et$ $al$, \ are illustrated by the black
points in Fig. 1. The continuous curve in the figure is determined
by a fit \ to the measured points after also adding two physical
conditions: a) \ The transition temperature in the limit $\nu->1$
should vanish because the density of the Skyrmion gas tends to
zero \cite{barret}, and b) To assume, in a first approximation, a
Skyrmion-anti-Skyrmion symmetry around $\nu=1$. \ The even
polynomial of the filling factor deviation $X=\nu-1,$\ employed
for the fit, \ included up\ to the sixth powers of $X$. \ \ It
should be noticed that the curves for the liquid to solid phase of
the Skyrmions estimated from the specific heat measurement of
Bayot $et$ $al$ \cite{bayot} \ do not show the  particle and
antiparticle symmetry. \ However, taking into account that the
vanishing of the critical temperature $T_{m}$ for $\nu=1$ should
be valid in any case (See, eg. Ref. \onlinecite{brey}), the
qualitative argument below will not be affected by this
simplifying assumption in the first stage of the analysis. \ The
qualitative behavior shown in Fig. 1 (apart from the above
mentioned asymmetry)\ is also shown by the measurements in Ref.
\onlinecite{bayot}. \ It will be seen that the analysis given here
will supply a possible mechanism for the lack of symmetry around
$\nu=1$.\ \

The behavior\ of $T_{m}$ in Fig.1\ can be studied by considering eqn. (1). The
derivative of $T_{m}$\ is depicted in Fig 2. Firstly, it may be noted that for
$X<0$ (the region in which the experimental points of Gervais $et$ $al$ were
measured) there exists a negative value of $X=X_{c}$ for which $\Delta M$
should be zero. This is so, because we expect physically \ that $S_{solid}$
$<S_{liquid}$. Defining now $\Delta S$=$S_{solid}-S_{liquid}=S_{s}-S_{l}$ (and
also $\Delta M=M_{s}-M_{l}$) it follows that $\Delta S$ is non vanishing and
negative. Then, the magnetization \ of the liquid \ at $X=X_{c}$ , is
identical to that of the solid at the zero of \ $T_{m}^{\prime}$ for $X=X_{c}%
$. \ \ \ Over $X_{c}$ up to $X=0$ the quantity $\Delta M$ should become
positive, since from thermodynamics \ $\Delta M/\Delta S<0,$ hence $\Delta M$
and $\Delta S$ must have the opposite sign \ for $X>X_{c}$ . Similarly, for
the region $X<X_{c},$ the magnetization change should acquire negative values.

The filling factor values for which the internal energies of the crystal and
the liquid coincide can also be estimated by considering the relation
\begin{equation}
\frac{\partial}{\partial\nu}\ T_{m}(\nu)=-\frac{T_{m}(\nu)}{\nu},
\end{equation}
which should be obeyed at those points \cite{leamarch}. This
equation has the following solutions$:X=-0.104$ and $X=0$, lying
in the interval defined by $X=-0.16$ (corresponding to the lowest
measured filling factor) and $X=0$. \ It is of interest that the
equal energy filling factors are very close to the values for
which the magnetization of the \ crystal and the liquid are also
coinciding. \ The equality of the internal energy at $X=0$, that
is at $\nu=1,$ is an expected result since the Skyrmion density
vanishes at $\nu=1.$

\ In addition, the inset of Fig. 3 of Gervais $et$ $al$ , as these workers
stress, \ suggests \ that there is a critical filling factor in the range
\ $\nu\approx0.80-0.83$ \ where a quantum phase transition exists. \ Following
the treatment of reference March, Suzuki and Parrinello\cite{marsupar}, it is
then relevant to ask what is the order parameter of this transition. We
propose the low frequency \ shear modulus, say $G$ , of the Skyrmion
\ crystal, \ as a candidate. If the quantum phase transition \ remains first
order, in analogy with (1) for the melting curve, then it follows \ from this
equation \ that, since \ \ the Third Law of Thermodynamics requires $\Delta
S->0$, that \ the melting curve \ has a positively infinite \ slope \ at the
critical filling factor \ near \ $0.8$. \ This assumes the unlikely event that
\ $\Delta M->0$ \ such that \ $\Delta M/\Delta S$ \ remains \ finite \ does
not occur. Also, such a first order transition would imply that the low
frequency shear modulus \ of the Skyrmion crystal \ goes discontinuously to
zero \ at the quantum phase transition.

\begin{figure}[h]
\includegraphics[width=3.0in]{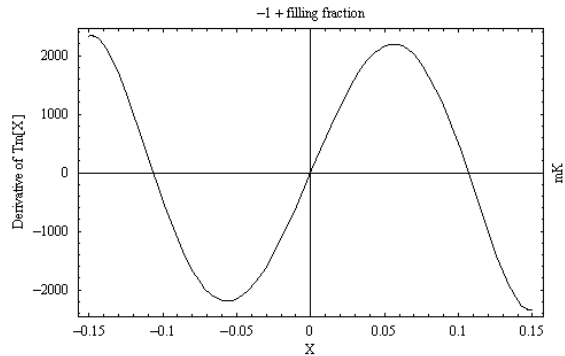}\caption{The derivative of the
estimated curve for $T_{m}(\nu)$ showing the various change of signs of
$\Delta M/\Delta S$ as indicated by Eq. (1).}%
\label{march}%
\end{figure}

We next advance an explanation for the form of the estimated solid-liquid
phase transition curve of an assembly of Skyrmions. \ The picture to be
exposed incorporates concepts coming from the \ 'anyon' type of description of
the \ QHE state near $\nu=1\cite{leamarch, kivelson, cabdann}$. We expect that
the qualitative explanation given can help to make contact between 'anyon'
\ descriptions and the transition \ between liquid and crystal Skyrmion phases.

Let us consider the region $X<0$\ near $\nu=1$ and first assume
that $X$ \ is close to zero. That is, \ the \ density of \
Skyrmion excitations $N$ is low and naturally they will tend to
crystallize \cite{brey}. \ This situation is illustrated at the
left hand side in Fig. 3 , \ which shows a crystalline arrangement
of Skyrmions. \ For \ $\nu<1$ \ the Skyrmions should really be of
the anti-Skyrmion kind having a positive net charge. This charge
distribution is indicated in the figure by the plus sign. \ The
curly arrows lying inside each Skyrmion picture signal the flow of
diamagnetic Hall currents, giving rise to the magnetic moment of \
the assumed almost isolated Skyrmion. This magnetic moment is
depicted as the white vertical arrow over each Skyrmion picture. \
Since, the temperature is non-vanishing and is also close to the
critical value $T_{m}$, the excitations will also have a kinetic
motion. This should produce a mean increment of the magnetic
moment per particle. \ \ It is clear that this increment will be
always a diamagnetic one: that is, opposite to the external
magnetic field $B$. \ This contribution is shown as a black arrow
below each of the particle symbols. \ \ In the crystal phase (left
hand side) it seems that the crystal potential could drastically
restrain the realization of \ large circular motion of the center
of mass of each Skyrmion at thermal velocities, around the crystal
position . Then, the diamagnetic orbital moment addition to each
Skyrmion $m_{S,o-s},$ can be expected to be reduced in absolute
value by the crystal potential. \newline

\ \ \ \ The liquid phase state of the Skyrmion gas is \ illustrated in the
right hand side of Fig. 3. \ The \ same conventions for the symbols are used
and the temperature is also assumed to be equal to $T_{m}$ but slightly above
the phase transition. \ In this case, since the gas is very diluted in the
considered situation, the Skyrmion proper magnetic moments can be expected to
be coincident in value with the ones in the crystal phase. \ The situation
seems to be different for the orbital component of the magnetic moment created
by the thermal fluctuations.\ In this case, the constraining action of the
crystal is not present and the \ Skyrmions have the chance of orbiting with
similar thermal velocities but in greater circles with appreciably more
degrees of freedom than their solid counterparts. Thus, an increase in
magnitude of the orbital magnetic moment contribution $m_{S,o-l}$ with respect
to its solid counterpart $m_{S,o-s}$, should be expected in the liquid phase.

\begin{figure}[h]
\includegraphics[width=3.0in]{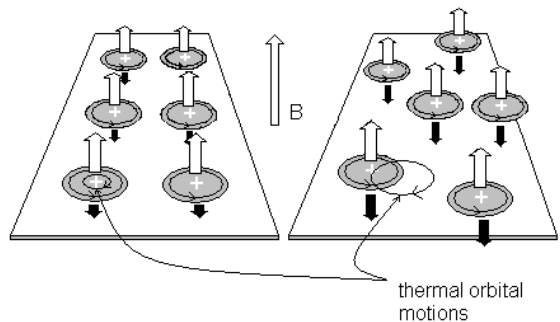}\caption{Diagrams representing the
Skyrmion crystal (left hand side) and the Skyrmion liquid (right hand side).
The white vertical arrows represent the individual magnetic moments of each
particle due to its internal (Hall) motion. The black vertical black ones
indicate the contribution to the magnetic moments coming from the thermally
induced orbital motion of the center of mass. \ The less constrained orbital
movement \ in the liquid phase can be expected to increase the thermally
induced orbital components. This is illustrated by the longer black arrows in
the liquid diagram.}%
\label{march}%
\end{figure}

Therefore, in the region \ $X<0$ \ down to $X_{c}$ \ it is to be expected that \ %
\begin{align*}
M_{s}-M_{l}  &  =N\text{ }(m_{S}+m_{S\text{,}o-s})-N(m_{S}+m_{S,o-l}),\\
&  =N\text{ }(-m_{S\text{,}o-l}+m_{S\text{,}o-s}),\\
\Delta M  &  =M_{s}-M_{l}>0.
\end{align*}
This is the same conclusion that the estimated phase diagram predicts,
according to eqn. \ (1).

\ \ As the density of anti-Skyrmions $N$ \ grows as \ $X$ diminishes
approaching to $X_{c}$ in the liquid phase, \ the scattering among the
Skyrmions \ becomes more intense. This effect can be expected to inhibit the
realization of large circular motions of the center of mass of orbiting
anti-Skyrmions, leading to a decrease of the magnitude of $\ m_{S\text{,}o-l}$
down to equalize it to $m_{S\text{,}o-s}$ at $X=X_{c}.$ \ Further, as the
density increases even more, the scattering plus overlapping effects of the
Skyrmions in the liquid phase makes natural a further reduction of the net
magnetization of the liquid with respect to the value in the more organized
solid arrangement, allowing steady currents and magnetic moments. \ This
situation is in accord with the property $\Delta M<0$\ following from the
estimated critical curve for $X<X_{c}<0.$ \newline

Let us propose below a qualitative explanation for the
Skyrmion-anti-Skyrmion observed asymmetry in the heat capacity
measurements in Ref. \onlinecite{bayot}.\ \ For this purpose we
notice that the orbital magnetic moment generated by the thermal
motion is not changing its sign when the filling factor\ passes
across the $\nu=1$ value. \ In addition it can be also underlined
that the \ Skyrmion has a rotating internal motion. These
circumstances are illustrated in Fig. 4. Therefore, such
rotational movements inside the crystal should generate shear
forces tending to distort it in some measure, as illustrated in
the left hand side of Fig. 4. But, after taking \ into account
that for \ $\nu>1$, both, the internal Skyrmion rotational motion
\ and the thermal one coincide in sense, and that, on the
contrary, for $\nu<1$ \ these motions are opposed, \ it follows
that the shear forces distorting the Skyrmion crystal should be
higher for $\nu>1$. But, the \ crystal melting is strongly
determined by the resistance to shear deformations, and then, the
$\nu>1$ crystal should be in a state closer to an instability
under shear deformations. Thus, it seems natural to expect that
the melting \ in the $\nu<1$ region will require higher
temperatures to be attained, as experimentally
observed\cite{bayot}.\ \ \

\begin{figure}[h]
\includegraphics[width=3.0in]{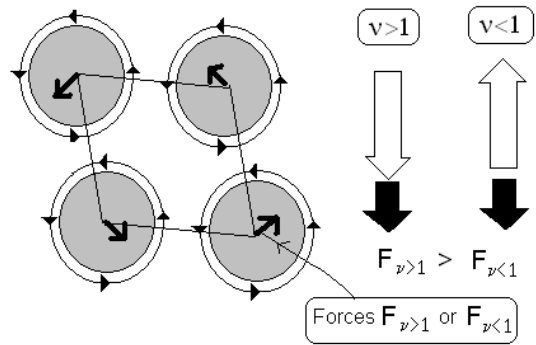}\caption{Diagrams illustrating the the
appearance of shear forces in a crystal formed by identical rotating bodies
(left hand side). The right hand side shows the two ways of combinations of
the internal \ rotations due to Hall (Chern-Simons) current and charge
densities constituting the Skyrmions, \ and the thermally induced diamagnetic
rotation of the center of mass of the excitation. \ For $\nu>1$ they are in
the same sense, and for $\nu<1$ \ they are opposed. Thus, for $\nu>1$ the
crystal should be expected to be more deformed.}%
\label{march}%
\end{figure}

Let us remark that in contrast to the behavior near $\nu=1$, where experiment
strongly points to the ground-state being a Skyrmion crystal \cite{barret,
gervais}, there is accord between experimental groups that a Skyrmion crystal
\ does not form \ the \ ground state \ of the two dimensional electron
assembly near $\nu=3$, where \ a Wigner crystal seems the most likely
candidate \ from existing experiments. \ \ Therefore, \ the whole situation
around integral filling factors seems to reinforce the view advanced in
\cite{cabdann, cabclarmarch} that the main ground state at exactly integer
filling factors \ shows an electromagnetic Chern-Simons (equivalent to a Hall
one\cite{cabdann}) response that either:

a) \ Transforms the set of impurities, when they exist in dirty samples, in
\ a kind of "charge reservoir", which accepts or releases electrons \ above or
below $\nu=1,$ respectively.

b) \ Or, in rough terms, \ generates "artificial impurities" localizing
charges, as Skyrmions or anti-Skyrmion excitations near $\nu=1$ and
'$electron$' or '$hole$' like Wigner crystals near $\nu=3$, again when $\nu$
is above or below the corresponding filling factor,\ respectively.

\ \ The coexistence of all these crystal or liquid structures for
magnetic fields well inside plateaus is a supporting experimental
fact for this view, as stated in \cite{cabdann, cabclarmarch}. The
demonstration of the validity of this property \ for realistic
planar samples is under consideration.\ In Ref.
\onlinecite{cabdann} it was only shown for the simpler case of a
superlattice of \ planar samples.

To conclude, near $\nu=1$ \ the measured phase diagram \ given by Gervais $et$
$al$ has been \ interpreted \ using the thermodynamic relation (\ref{marchlea}%
) \ and making the plausible assumption \ that $S_{s}<S_{l}$. The measured
melting curve requires various change of signs of the magnetization
\ difference $\Delta M$ \ in the region $\nu=0.8-1.2$ \ A qualitative
interpretation of \ the nature of these changes \ is advanced by considering
the internal structure of the Skyrmions and their thermal motions. \ The
observed asymmetry of the melting into a Skyrmion liquid phase above and below
$\nu=1$ is proposed to be produced by the diamagnetic moment created by the
thermal motion, which reduces or increases the magnitude of the internal
magnetic moments of the Skyrmions and anti- Skyrmions, respectively.
Therefore, an explanation\ of the observed properties of the Skyrme
excitations near $\nu=1$ is proposed and based on an electromagnetic and
'anyon' description \cite{kivelson, leamarch, cabdann}. However, further
\ work, both experiment and theory, \ is required on this potentially
important matter.

\ \

\section{Acknowledgments}

N.H.M and A.C.M wish to acknowledge that their contribution to
this article was brought to fruition \ during a stay at the AS
ICTP, Trieste in 2005. This stay was made possible by Prof. V.E.
Kravtsov, to whom they are grateful for generous hospitality and
the very stimulating environment. The support of \ FONDECYT,
Grants 1020829 and 7020829, for the activity of \ F.C. is also
very much recognized.

\end{document}